# Node-neighbor subnetworks and $H_k$-core decomposition

Dinghua Shi[1], Yang Zhao[2] and Guanrong Chen[3]
[1]Shanghai University, Shanghai 200444, China, shidh2012@sina.com
[2]Fudan University, Shanghai 200433, China, 21110720040@m.fudan.edu.cn
[3]City University of Hong Kong, Hong Kong 999077, China, eegchen@cityu.edu.hk

**Abstract**

The network homology $H_k$-core decomposition proposed in this article is similar to the *k*-core decomposition based on node degrees of the network. The C. elegans neural network and the cat cortical network are used as examples to reveal the symmetry of the deep structures of such networks. First, based on the concept of neighborhood in mathematics, some new concepts are introduced, including such as node-neighbor subnetwork and Betti numbers of the neighbor subnetwork, among others. Then, the Betti numbers $(\beta_0, \beta_1, \ldots, \beta_k)_i$ of the neighbor subnetwork of node *i* are computed, which are used to perform $H_k$-core decomposition of the network homology. The construction process is as follows: the initial network is referred to as the $H_0$-core; the $H_1$-core is obtained from the $H_0$-core by deleting the nodes where $(\beta_0=1, \beta_1=0, \ldots, \beta_k=0)_i$; the $H_2$-core is obtained from the $H_1$-core by deleting the nodes or edges where $(\beta_0>1, \beta_1 \geq 0, \beta_2=0, \ldots, \beta_k=0)_i$ in which $\beta_1=0$ means deletion of nodes and $\beta_1>0$ means removal of edges connected to the $\beta_0$-related branches in the subnetwork; the $H_3$-core is obtained from the $H_2$-core by deleting the nodes where $(\beta_0 \leq \beta_1, \beta_2=0, \ldots, \beta_k=0)_i$ or by retaining the nodes where the Betti numbers satisfy $(\beta_{2 \text{ or } k} >0)_i$, and so on. Throughout the process, the index of node involved in deleting edge needs to be updated in every step. The $H_k$-core decomposition is easy to implement in parallel. It has a wide range of applications in many fields such as network science, data science, computational topology, and artificial intelligence. In this article, we also show how to use it to simplify homology calculation, e.g. for the C. elegans neural network with ($V$=297, $E$=2148; $\beta_0$=1, $\beta_1$=139, $\beta_2$=121, $\beta_3$=4), whereas the results of decomposition are the $H_1$-core with (226, 1723; 1, 139, 121, 4), the $H_2$-core with (154, 1184; 1, 0, 121, 4), and the $H_3$-core with (16, 68; 1, 0, 0, 4). Thus, the simplexes consisting of four highest-order cavities in the $H_3$-core subnetwork can also be directly obtained.

**Keywords:** node-neighbor subnetwork, characteristic number, boundary matrix, Betti number, $H_k$-core decomposition, homology calculation.

## 1 Introduction

The introduction of small-world networks [1] and scale-free networks [2] at the turn of the century ushered in a new science of networks. The collective dynamics of networks broke through the limitations of traditional graph theory research and combined dynamics with graph structures [3], making network design an important and interesting research topic [4]. The subsequent investigations of totally-homogeneous networks [5] and simplicial networks [6] emphasized that the view of network structure should be shifted from stars to cycles [7]. This shift reflects the transition from lower-order to higher-order networks [8], thereby opening up a new research direction for using homology theory in algebraic topology to study network science.

Homology is a tool in algebraic topology [9] that describes the graphic information contained in different dimensional spaces. Specifically, 0-dimensional homology studies network connected branches, 1-dimensional homology studies cycles of simplicial networks (or complexes), 2-dimensional homology studies cavities surrounded by triangles [10], and so on. The advances of the big data era, especially the emergence of higher-dimensional data (e, g., the point cloud data [11] in a metric space), have spawned a new discipline of topological data analysis (TDA), which reflects the intersection of data science, network science and computational topology [12]. It is now clear that the observed data shape can only be considered a true representation of the original data if the topological features appear continuously at multiple spatiotemporal scales; otherwise, it may just be numerical errors caused by sampling and noise.

Persistent homology [13] is a method for studying the homology changes in a parameterized space. First, a series of distances are given to the point cloud data, for example, embedding the network in a metric space, or using the inverse of the edge weight or the shortest path between nodes to characterize the distance. The constructed nested simplicial complexes are called a filtration [14,15]. Then, the homology of every dimension of each simplicial complex in the filter is calculated. In addition to determining the birth time and death time of the Betti barcodes, the more important task about homology calculation is to find the persistent Betti barcodes. However, it is quite difficult to calculate the simplices representing the strip diagram (forming the cavities) [15,16].

In addition to the above-mentioned perspective of introducing homology from TDA, another way to introduce homology is to explore the influence of network structure on dynamics from the perspective of network dynamics. A case in point is network synchronization, which is related to the smallest non-zero eigenvalue of the network Laplacian matrix (denoted as $\lambda_2$); the larger the $\lambda_2$ is, the easier it is for the network to achieve synchronization [3]. In this consideration, it is natural to ask: Under the same network size, what kind of network structure has the largest $\lambda_2$? We found that it is a fully-homogeneous network structure with perfect symmetry [5]. Fully-homogeneous structures such as cycles, cliques and cavities are the cornerstones of higher-order networks and homology groups [5].

Recall that the cliques in graph theory are called simplices in algebraic topology. For example, a node is a 0-order clique, an edge is a 1-order clique, a triangle is a 2-order clique, a tetrahedron is a 3-order clique, and so on. It is worth noting that, in practical problems, some triangles are not necessarily 2-order simplices [17]. Note that there are many ways to introduce simplices, the simplicial complex composed of all simplices in the network must satisfy the inclusion relation. Specifically, if $\sigma$ is a simplex in the network, then the lower-order $\tau \subseteq \sigma$ must also be a simplex, that is, $\sigma \in K \wedge \tau \subseteq \sigma \rightarrow \tau \in K$. The number of $k$-order simplices in the network is denoted by $m_k$, with which the first topological invariant of the network is given by $\chi = m_0 - m_1 + m_2 - \cdots$, called the Euler characteristic [16].

In the study of star structure in a network, the row sum of the adjacency matrix is used to represent the node degree, but in order to study the cycle structure, new mathematical tools are needed. First, all $k$-order simplexes $\{\sigma_i\}$ form an $m_k$-dimensional vector space, denoted $C_k$. Vector addition uses the set symmetric difference operation, and the number field of multiplication can be a binary field or a real number field. In addition to using the common vector space, distance space and topological space [18] can also be used.

Vector spaces composed of adjacent-order simplices can be connected through mapping. For example, the boundary operator $\partial_k: C_k \rightarrow C_{k-1}$ can be introduced [5,19]. With this boundary operator, some subspaces of the vector space can be further studied. In the real number field, the chain in $C_k$ is defined as $l_k = \Sigma g_i \sigma_{ik}$, where $g_i$ is an integer; in the binary field, $g_i \equiv 1$. If $\partial_k l_k = 0$, then $l_k$ is a cycle in $C_k$. Obviously, $\partial_{k+1} l_{k+1}$ is also a cycle in $C_k$, called an exact cycle [20]. The subspace composed of all cycles in $C_k$ is denoted as $\ker(\partial_k)$ or $Z_k$, which is called the kernel space; the subspace composed of all exact cycles in $C_k$ is denoted as $\text{im}(\partial_{k+1})$ or $Y_k$, which is called the image space. Obviously, $Y_k \subseteq Z_k$, so the quotient space $Z_k/Y_k$ can be studied, which is the $k$-order homology group denoted as $H_k = \ker(\partial_k)/\text{im}(\partial_{k+1})$. There are naturally different ways to define other homology groups [21].

In order to facilitate the study of boundary operators, they are usually characterized by a (oriented) boundary matrix $B_{[k]}$ [19]. Sometimes it is easier to consider the boundary matrix $B_k$ [16] defined on a binary field. Denote the rank of the boundary matrix $B_k$ by $r_k$. Then, the second topological invariant of the network $\beta_k = m_k - r_k - r_{k+1}$, referred to as the $k$-order Betti number [16]. The 0-order Betti number $\beta_0$ is equal to the number of connected branches, and the $k$-order Betti number $\beta_k$ is the rank of the $k$-order homology group or the number of cavities in the network.

In this way, all networks, regardless of their sizes, have a triplet $(m_k, r_k, \beta_k)$, for which a calculation method is presented in [16]. The original network size is relatively large, so its calculation is fairly difficult. Nevertheless, there is only one (the original) network whereas its node neighbor subnetwork is relatively small. Therefore, although the network size is large, the calculation is relatively easy and can be implemented in parallel. In addition to determining the number of cavities, $\beta_k$, the calculation of the homology group is more difficult, which needs to find the minimum number of simplices that form the cavities. In [16], it is proposed to use an optimization method to find the shortest length (minimum number) of the cavities, and in [15] it was proposed to use a spanning-tree method to find all cavities of the same order. Here, in this paper, it is proposed to use the information of the triplet $(m_k, r_k, \beta_k)$ of the node neighbor subnetwork to study the $H_k$-core decomposition of the original network, and then use the decomposition information and the new node index to simplify the calculation of various cavities and find the highest-order cavities.

The rest of the paper is organized as follows: The Methods section introduces the neighbor subnetwork, a new node index, and the homology $H_k$-core decomposition through simple explanatory examples. The Results (and Applications) section applies the $H_k$-core decomposition to two real networks, the C. elegans neural network and the cat cortical network, describes the $H_k$-core decomposition process of network homology, and introduces a new method for finding the highest-order cavities in the network with the help of the new node index. Finally, the Discussion section discusses other possible applications of the $H_k$-core decomposition and some related issues.

## 2 Method

**Neighbor Subnetwork**

The concept of neighbor subnetwork is introduced. It is small in size but it outlines the original whole network. In the neighbor subnetwork, in addition to the usual neighbor number $n_i$, characteristic number $\chi_i = 1 - \chi^\wedge$ and Betti number $(\beta_k)_i$, it may also have special node degrees such as isolate node degree ($d_{\min}=0$) and central node degree ($d_{\max}=n_i-1$), see Fig. 1

In Fig. 1(a), the red node and red edges are tree to be deleted; black node and edges constitute a subnetwork, with characteristic number denoted by $\chi^\wedge$. For example, in subfigure (1) the black subgraph is a neighbor subnetwork with 3 nodes and 1 edge, where the red-circle node is an isolated node, with characteristic number $\chi^\wedge=3-1=2$ and Betti numbers $\beta_0=2$, $\beta_1=0$. The node to be deleted has characteristic number $\chi_i=1-\chi^\wedge=-1$. The subfigure (4) has 4 nodes and 3 edges, where the green node is the central node, with characteristic number $\chi^\wedge=4-3=1$ and Betti numbers $\beta_0=1$, $\beta_1=0$. The node to be deleted has characteristic number $\chi_i=1-\chi^\wedge=0$, where the Betti number $\beta_0$ is equal to the number of branches of the subnetwork, in which the first subnetwork has 2 branches, 1 isolated node and 1 branch without cycle.

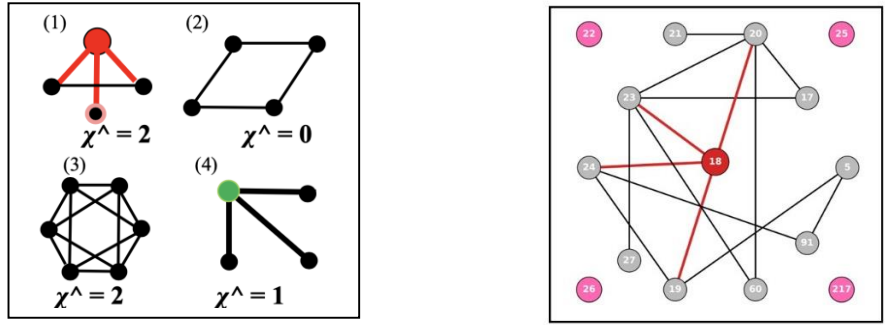

Fig. 1 (a) Neighbor subnetwork examples; (b) C. elegans neighbor subnetwork

Figure 1 (b) shows the neighbor subnetwork of node 16 in the C. elegans neural network [22], which consists of 15 nodes, 15 edges and 4 triangles, denoted ▲ below. It has Betti numbers $\beta_0=5$, $\beta_1=1$. Moreover, it has 5 branches: 4 are simply isolate nodes 22, 25, 26, 217 (wherein they are marked red), while another branch has 11 nodes with $\beta_1=1$. Furthermore, the other 7 nodes constitute a rather complicated structure without $\beta_1$ but containing 4 triangles: ▲17-20-23, ▲18-19-24, ▲18-20-23 and ▲20-23-60, as well as edge 20-21 and edge 23-27.

The central node and the Betti number of the neighbor subnetwork have the following relationship.

**Proposition 1** If the neighbor subnetwork of node $i$ has a central node, then the neighbor subnetwork has Betti numbers $(1, 0, 0, \ldots, 0)_i$.

It is obvious that if the neighbor subnetwork of node $i$ has a central node, then the subnetwork is connected, so the neighbor subnetwork has Betti number $\beta_0=1$. Since any 1-order cycle of the subnetwork with length longer than 4 cannot have central node, which will also be equivalent to some triangle formed by the central node and an edge in the cycle, thus the Betti number of the subnetwork is $\beta_1=0$. Similarly, any $k$-order cycle of the subnetwork cannot contain central nodes, which will be equivalent to some $(k+1)$-order simplex formed by the central node and any $k$-order simplex in the $(k+1)$-order simplex, therefore the Betti number of the subnetwork is $\beta_k=0$. Consequently, the neighbor subnetwork of node $i$ has Betti numbers $(1, 0, 0, \ldots, 0)_i$.

**New Node Index**

Here, the new node index is defined to be a triplet of parameters: number of neighbors, Betti number, and the characteristic number of the network consisting of all simplices to be deleted ($\chi_i=1-m_0^\wedge+m_1^\wedge-m_2^\wedge+\cdots$). For instance, the first subnetwork in Fig. 1(a) has $\chi_i=1-3+1=1-\chi^\wedge=-1$. Thus, node $i$ has a new index $\{n_i, (\beta_k)_i, \chi_i\}$. If necessary, the index may also include the maximum degree (of the central node) and the minimum degree (of the isolated node).

In Fig. 1(a), subnetwork (1) has a new index $\{3, d_{min}=0, (2,0), -1\}$; subnetwork (2) has 4 nodes, 4 edges, characteristic number $\chi^{\wedge}=4-4=0$, Betti numbers $\beta_0=1, \beta_1=1$, so its index is $\{4, (1,1), 1\}$; subnetwork (3) has 6 nodes, 12 edges, 8 triangles, characteristic number $\chi^{\wedge}=6-12+8=2$, Betti numbers $\beta_0=1, \beta_1=0, \beta_2=1$, so its index is $\{6, (1,0,1), -1\}$; subnetwork (4) has index $\{4, d_{max}=3, (1,0), 0\}$.

**Illustrative Example**

Figure 2 is a sample network discussed in [16], which has 14 nodes, 26 edges, 13 triangles and 1 tetrahedron. It has characteristic number $\chi=14-26+13-1=0$, Betti numbers $\beta_0=1, \beta_1=2, \beta_2=1$, and therefore the new node index $\{n_i, (\beta_k)_i, \chi_i\}$ indicates: nodes 1, 2: $\{4, d_{max}=3, (1,0,0), 0\}$, 3: $\{5, d_{min}=0, (3,0,0), -2\}$, 4: $\{3, d_{max}=2, (1,0,0), 0\}$, 5: $\{3, (2,0,0), -1\}$, 6: $\{3, d_{min}=0, (3,0,0), -2\}$, 7, 8: $\{2, d_{min}=0, (2,0,0), -1\}$, 9, 14: $\{5, d_{min}=0, (2,1,0), 0\}$ participated in both 1- and 2-order cavities, 10, 11, 12, 13: $\{4, (1,1,0), 1\}$. These new node indices provide important information about the network topology, laying a basis for the network $H_k$-core decomposition.

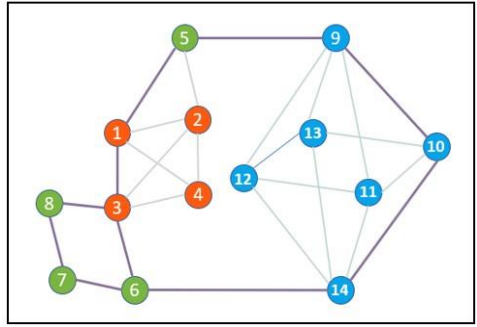

Fig. 2 A sample network

**Proposition 2** If the largest non-zero Betti number of a network is $\beta_k$, then the largest non-zero Betti number of the subnetwork will not exceed $\beta_k$.

This proposition can be used to reduce the amount of calculation of the new index of the node by truncating it to $k$.

Note that the largest $k$-order cavity of the subnetwork will not disappear in the case of adding no internal connections. When adding connections with external nodes and edges, they can only be replaced by cavities of the same or higher order due to the cycle equivalence relationship. Therefore, the largest non-zero Betti number of the subnetwork will not exceed the largest non-zero Betti number of the original subnetwork.

**Homology $H_k$-core Decomposition**

Following the $k$-core decomposition of node degrees [23], the $H_k$-core decomposition of network homology also deletes nodes and edges. First, it is necessary to delete redundant nodes and edges for homology, and then delete all nodes and edges that generate 1-, 2-, …, $(k-1)$-order cavities one by one.

**Proposition 3** If the Betti numbers of the neighbor subnetwork of node $i$ are $(\beta_0, 0, …, 0)_i$, then when $\beta_0=1$, deleting node $i$ does not change the network topology; that is, the characteristic number and Betti numbers of the original network can be maintained after deleting the nodes and edges; when $\beta_0>1$, deleting node $i$ changes the Betti number of the network, so that $\beta_1$ decreases by $\beta_0-1$.

When $\beta_0=1$, the neighbor subnetwork of node $i$ has Betti numbers $(1, 0, \ldots, 0)_i$ therefore node $i$ has characteristic number 0, so deleting node $i$ does not change the characteristic number of the subnetwork.

When $\beta_0=1$, deleting node $i$ and its edges as well as subnetwork-related elements always form some simplices of different orders, so deleting node $i$ will not affect the Betti number of the network. When $\beta_0>1$, arbitrarily selecting a branch to delete the edge between node $i$ and a branch node will only reduce the Betti number of the subnetwork by 1. After selecting $\beta_0-1$ branches, the Betti number of the neighbor subnetwork of node $i$ is $(1, 0, \ldots, 0)_i$.

**Proposition 4** Deleting the edges connected to nodes without $\beta_k$ branch of the neighbor subnetwork of node $i$ will reduce the value $\beta_k$.

Consider a branch without $\beta_1$ as an example, it can be generalized. If there is an isolated node without $\beta_1$ branch, then the value $\beta_1$ of the network will decrease by 1 after deleting an edge connected to the isolated node, so the conclusion is established. If it is not an isolated node, then the network topology will remain unchanged after deleting the edges, but the number of nodes contained in the branch will decrease. This process will continue until an isolated node appears in the end.

**Homology $H_k$-core Decomposition Algorithm**

The original network is denoted as the $H_0$-core. Compute the triplet $(m_k, r_k, \beta_k)$ and determine the value $k$ of the maximum non-zero $\beta_k$. so as to obtain a new node index $\{n_i, (\beta_k)_i, \chi_i\}$, including the $d_{\max}$ value of the central node and the value $d_{\min}$ of the isolated nodes when needed.

**Step 1** is deleting nodes: To obtain the $H_1$-core of the network, it needs to delete redundant nodes and edges for homology. Specifically, by propositions, delete the nodes $i$ with Betti numbers $(\beta_0=1, \beta_1=0, \ldots, \beta_k=0)_i$ one by one, according to the node degrees from small to large. After each deletion, the index of node involved in deleting edge needs to be updated until there are no nodes with Betti numbers $(\beta_0=1, \beta_1=0, \ldots, \beta_k=0)_i$. In order to increase the processing speed of the algorithm, one can first calculate node $d_{\max}$ value in the subnetwork and delete the central nodes, and then calculate the Betti numbers and delete the nodes with $(\beta_0=1, \beta_1=0, \ldots, \beta_k=0)_i$. After completion, calculate the triplets $(m_k, r_k, \beta_k)$ and $\{n_i, (\beta_k)_i, \chi_i\}$ of the $H_1$-core.

**Step 2** is deleting nodes and edges: In order to obtain the $H_2$-core, it is necessary to delete the nodes or edges that generate the network Betti number $\beta_1$ in the network $H_1$-core. According to the propositions, following the node degrees from small to large, delete the nodes $i$ with Betti numbers $(\beta_0>1, \beta_2=0, \ldots, \beta_k=0)_i$ one by one (in order to improve the processing speed, one can also keep the subnetwork composed of nodes $i$ with Betti numbers $(\beta_{1 \text{ or } k}>0)_i$, but it needs to check that the network $\beta_k$ $(k>1)$ remains unchanged). Likewise, it needs to be updated after each deletion. In addition, it is necessary to delete the newly appearing nodes $i$ with Betti numbers $(\beta_0=1, \beta_1=0, \ldots, \beta_k=0)_i$ one by one, until there are no nodes with Betti numbers $(\beta_0>1, \beta_2=0, \ldots, \beta_k=0)_i$. At this time, the network $\beta_1$ has been reduced by $\Sigma(\beta_0-1)_i$. If $\beta_1$ is still greater than 0, then delete the edges connecting without $\beta_1$ branches in the neighbor subnetworks for the nodes $i$ containing $(\beta_0>1, \beta_1>0, \beta_2=0, \ldots, \beta_k=0)_i$. Each deletion also needs to be updated until the network Betti number $\beta_1=0$. After completion, calculate the $(m_k, r_k, \beta_k)$ and $\{n_i, (\beta_k)_i, \chi_i\}$ of the $H_2$-core.

**Step 3** is deleting edges: In order to obtain the $H_3$-core, it is necessary to delete the edges with positive characteristic numbers in the $H_2$-core. Delete the nodes $i$ with Betti numbers $(\beta_0 \leq \beta_1, \beta_2=0, \ldots, \beta_k=0)_i$ one by one, according to the degrees ranging from small to large (or keep the nodes $i$ with Betti numbers $(\beta_{2\ or\ k}>0)_i$, but it needs to check that the network $\beta_k$ ($k>2$) remains unchanged). Then, update it after each deletion. In addition, it is necessary to delete the newly appeared nodes $i$ with Betti numbers $(\beta_0=1, \beta_1=0, \ldots, \beta_k=0)_i$ one by one. If $\beta_2$ is still greater than 0, then, delete the edges connecting the nodes $i$ with $(\beta_0 \leq \beta_1, \beta_2>0, \beta_3=0, \ldots, \beta_k=0)_i$ to the neighboring subnetwork with $\beta_2$ branches to obtain the $H_3$-core. After completion, calculate the $(m_k, r_k, \beta_k)$ and $\{n_i, (\beta_k)_i, \chi_i\}$ of the $H_3$-core.

**Step 4** is deleting the subnetwork: In order to obtain a higher-order $H_k$-core, keep the subnetwork composed of nodes $i$ with Betti numbers $(\beta_{k-1\ or\ k}>0)_i$, but make sure that the network $\beta_k$ remains unchanged. Then, delete the newly appeared nodes $i$ with Betti numbers $(\beta_0=1, \beta_1=0, \ldots, \beta_k=0)_i$ one by one. Continue to delete the obtain a subnetwork but keep the subnetwork composed of nodes $i$ with $(\beta_{k-1\ or\ k}>0)_i$, until the $H_k$-core is obtained. After completion, calculate the $(m_k, r_k, \beta_k)$ and $\{n_i, (\beta_k)_i, \chi_i\}$ of the $H_k$-core.

The previous two steps discuss in detail the relationship between $\beta_0$ and $\beta_1$; after that, $\beta_2, \ldots, \beta_k$ can be similarly discussed.

**Illustrative Example**

Figure 2 is a sample network used to illustrate the $H_k$-core decomposition process.

Compute the triplet $(m_k, r_k, \beta_k)$ of the original network, namely the $H_0$-core, so as to determine the maximum non-zero $\beta_k$ for $k = 2$.

Number of simplices: $m_0=14$, $m_1=26$, $m_2=13$, $m_3=1$; ranks of boundary matrices: $r_0=0$, $r_1=13$, $r_2=11$, $r_3=1$; Betti numbers: $\beta_0=1$, $\beta_1=2$, $\beta_2=1$, $\beta_3=0$; characteristic number: $\chi=0$.

Step 1: Delete nodes (all nodes with zero characteristic numbers). First, delete node 4, {3, $d_{max}=2$, (1,0,0), 0}. Nodes 1, 2, 3 are updated to be 1, 2: {3, $d_{max}=2$, (1,0,0), 0}, 3: {4, $d_{min}=0$, (3,0,0), −2}. Then, delete node 2: {3, $d_{max}=2$, (1,0,0), 0}, update nodes 1, 3, 5 to be 1; {2, $d_{min}=0$, (2,0,0), −1}, 3: {3, $d_{min}=0$, (3,0,0), −2}, 5: {2, $d_{min}=0$, (2,0,0), −1}. After updating, the characteristic number of node 1 has been changed from 0 to be negative, hence cannot be deleted. The process is thus ended, yielding the triplet $(m_k, r_k, \beta_k)$ of the $H_1$-core as follows:

Number of simplices: $m_0=12$, $m_1=20$, $m_2=8$; ranks of boundary matrices: $r_0=0$, $r_1=11$, $r_2=7$; Betti numbers: $\beta_0=1$, $\beta_1=2$, $\beta_2=1$; characteristic number: $\chi=0$.

Step 2: Delete nodes (all nodes with negative characteristic numbers). First, delete node 1: {2, $d_{min}=0$, (2,0,0), − 1}. Then, after updating, delete node 5 in (1, 0, 0)$_i$. After updating, delete node 3: {2, $d_{min}=0$, (2,0,0), −1}. Then, after updating, delete node 8, 7, 6 in (1, 0, 0)$_i$ without the need of deleting edges, and the process ends; or, keep the subnetwork consisting of nodes

9, 10, 11, 12, 13, 14 with $(\beta_{1 \text{ or } k} > 0)_i$. The triplet $(m_k, r_k, \beta_k)$ of the $H_2$-core are obtained as follows:

Number of simplices: $m_0=6$, $m_1=12$, $m_2=8$; ranks of boundary matrices: $r_0=0$, $r_1=5$, $r_2=7$; Betti numbers: $\beta_0=1$, $\beta_1=0$, $\beta_2=1$; characteristic number: $\chi=2$.

The nodes of the sample network shown in Figure 2 are divided with different characteristic numbers: zero-number nodes: 4, 2; negative-number nodes: 1, 5, 3, 8, 7, 6; and positive-number nodes: 9, 10, 11, 12, 13, 14. To this end, deleting nodes with zero characteristic numbers yields the $H_1$-core; deleting nodes with negative characteristic numbers, 1, 5, 3, 8, 7, 6, or keep nodes with positive characteristic numbers, 9, 10, 11, 12, 13, 14, yields the $H_2$-core.

## 3 Results (and Applications)

In this section, the $H_k$-core decomposition is applied to two real networks, showing the network homology $H_k$-core decomposition process, and introduce a new computing method for finding highest-order cavities based on the node new index measure.

**Example 1** Homology $H_k$-core decomposition for C. elegans neural network.

The C. elegans neural network has a triplet $(m_k, r_k, \beta_k)$ in its $H_0$-core as follows [16], having the maximum nonzero $\beta_k$ with $k=3$.

Simplices: $m_0=297$, $m_1=2148$, $m_2=3241$, $m_3=2010$, $m_4=801$, $m_5=240$, $m_6=40$, $m_7=2$; ranks of boundary matrices: $r_0=0$, $r_1=296$, $r_2=1713$, $r_3=1407$, $r_4=599$, $r_5=202$, $r_6=38$, $r_7=2$; Betti numbers: $\beta_0=1$, $\beta_1=139$, $\beta_2=121$, $\beta_3=4$, $\beta_4=\cdots=\beta_7=0$; characteristic number: $\chi=-21$.

Step 1: Delete nodes (all nodes with zero characteristic numbers).

Delete nodes with a central point in the neighborhood, totally 50: 243, 244, 261, 286, 287, 288, 289, 290, 291, 292, 293, 294, 295, 296, 297, 77, 98, 137, 145, 155, 197, 222, 228, 230, 231, 232, 235, 238, 239, 250, 251, 252, 255, 259, 269, 270, 271, 272, 273, 274, 275, 276, 277, 278, 279, 280, 281, 282, 283, 284 (wherein, red node has degree 1, node 155: {10, $d_{\max}=8$, (1,0,0,0), 0} and node 284: {8, $d_{\max}=6$, (1,0,0,0), 0}, which became a central node after updating, therefore was deleted. Note that such nodes are shadowed below.)

Delete nodes with Betti numbers $(\beta_0=1, \beta_1=0, \beta_2=0, \beta_3=0)_i$, totally 21: 18, 30, 46, 127, 130, 142, 144, 153, 159, 176, 177, 183, 184, 193, 245, 248, 254, 256, 258, 262, 285 (wherein, node 177: {11, (1,1,0,0), 1} and node 258: {10, (1,1,0,0), 1}, which became node $(1, 0, 0, 0)_i$ after updating, therefore was deleted.)

Network $H_1$-core has a triplet $(m_k, r_k, \beta_k)$ as follows:

Simplices: $m_0=226$, $m_1=1723$, $m_2=2582$, $m_3=1598$, $m_4=670$, $m_5=215$, $m_6=39$, $m_7=2$; ranks of boundary matrices: $r_0=0$, $r_1=225$, $r_2=1359$, $r_3=1002$, $r_4=592$, $r_5=178$, $r_6=37$, $r_7=2$; Betti numbers: $\beta_0=1$, $\beta_1=139$, $\beta_2=121$, $\beta_3=4$, $\beta_4=\cdots=\beta_7=0$; characteristic number: $\chi=-21$.

Step 2: Delete nodes (all nodes with negative characteristic numbers).

Delete nodes with Betti numbers $(\beta_0>1, \beta_1=0, \beta_2=0, \beta_3=0)_i$, totally 64: 22, 25, 26, 28, 29, 33,

34, 37, 38, 40, 43, 52, 55, 56, 57, 61, 79, 80, 81, 82, 83, 84, 86, 89, 90, 93, 95, 97, 101, 105, 107, 122, 128, 129, 135, 141, 168, 170, 172, 174, 175, 186, 189, 209, 210, 213, 217, 223, 224, 226, 233, 234, 236, 237, 241, 247, 249, 260, 263, 264, 265, 266, 267, 268 (wherein, the originally non-negative ones 33, 57, 84, 86 were deleted after their changes due to updating, but the originally negative ones 54, 69, 113, 242 were kept after their changes due to updating.)

Delete new nodes with Betti numbers $(\beta_0=1, \beta_1=0, \beta_2=0, \beta_3=0)_i$, totally 6: 41, 54, 69, 113, 179, 242 (This is the same as the previous case where nodes with Betti numbers $(\beta_{1或k}>0)_i$ are first kept and then the nodes with newly appearing Betti numbers $(1, 0, 0, 0)_i$ are deleted after updating. In this case, the network has (node-edge) number (156, 1218) and Betti numbers (1, 17, 121, 4), whereas it is impossible to delete all 1-order cavities by deleting only nodes, for which deleting edges is necessary.)

Delete edges in $(\beta_0>1, \beta_1>0, \beta_2=0, \beta_3=0)_i$ with isolated nodes, totally 17: 16-23, 19-73, 19-75, 20-47, 21-59, 21-66, 21-72, 23-72, 23-102, 32-50, 64-216, 73-96, 94-146, 138-229, 160-207, 187-191, 215- 240. (The neighbor subnetwork of node 16 is shown in Fig. 1(b), there are two branches after removing the red nodes and edges, where node 23 in the $\beta_0$ branch is just a representative of nodes 17, 20, 21, 23, 27, 60. Similarly, node 66 represents nodes 20, 63, 66 in the $\beta_0$ branch in the neighbor subnetwork of node 21, while node 22 has newly appeared Betti numbers $(1, 0, 0, 0)_i$ and node 23 has newly appeared Betti numbers $(2, 0, 0, 0)_i)$.

The triplet $(m_k, r_k, \beta_k)$ of the network $H_2$-core are as follows.

Simplices: $m_0=154$, $m_1=1184$, $m_2=2051$, $m_3=1317$, $m_4=573$, $m_5=193$, $m_6=36$, $m_7=2$; ranks of boundary matrices: $r_0=0$, $r_1=153$, $r_2=1031$, $r_3=899$, $r_4=414$, $r_5=159$, $r_6=34$, $r_7=2$; Betti numbers: $\beta_0=1$, $\beta_1=0$, $\beta_2=121$, $\beta_3=4$, $\beta_4=\cdots=\beta_7=0$; characteristic number: $\chi=118$.

Step 3 Delete nodes and edges (or, keep the subnetwork)

Delete nodes with Betti numbers $(\beta_0\leq\beta_1, \beta_2=0, \beta_3=0)_i$, totally129: 1, 2, 4, 5, 6, 8, 9, 10, 11, 12, 14, 15, 16, 17, 19, 21, 24, 27, 31, 32, 35, 36, 39, 42, 44, 45, 47, 48, 50, 53, 58, 59, 60, 62, 63, 64, 65, 66, 67, 68, 70, 71, 72, 73, 74, 75, 76, 78, 88, 91, 92, 94, 96, 99, 100, 102, 103, 104, 106, 108, 109, 110, 111, 112, 114, 115, 116, 117, 121, 123, 124, 125, 126, 131, 132, 133, 134, 136, 138, 139, 140, 143, 146, 147, 148, 149, 150, 151, 157, 160, 161, 165, 166, 167, 169, 178, 180, 181, 182, 187, 188, 190, 191, 192, 194, 196, 198, 199, 202, 204, 205, 206, 207, 208, 211, 212, 214, 215, 216, 218, 219, 220, 221, 225, 229, 240, 246, 253, 257.

Delete nodes with newly appeared Betti number $(\beta_0=1, \beta_1=0, \beta_2=0, \beta_3=0)_i$ after updating, totally 9: 7, 49, 51, 87, 120, 152, 156, 200, 203 (This is the same as the case where the newly appeared nodes with Betti numbers $(\beta_{2\ or\ k}>0)_i$, wherein 28 (27) nodes coming from the $H_0$-core ($H_1$-core), including node 179 (nodes 24 and 45) and 25 nodes with Betti numbers $(\beta_{2\ or\ k}>0)_i$ coming from the $H_2$-core. Delete nodes with newly appeared Betti numbers $(1, 0, 0, 0)_i$, totally 9: 7, 49, 51, 87, 120, 152, 156, 200, 203.

The rest nodes consist of $H_3$-core, totally 16: 3, 13, 85, 118, 119, 154, 158, 162, 163, 164, 167, 171, 173, 185, 195, 227.

The triplet $(m_k, r_k, \beta_k)$ of the network $H_3$-core are as follows.

Simplices: $m_0=16$, $m_1=68$, $m_2=113$, $m_3=68$, $m_4=4$; ranks of boundary matrices: $r_0=0$, $r_1=15$,

$r_2$=53, $r_3$=60, $r_4$=4; Betti numbers: $\beta_0$=1, $\beta_1$=0, $\beta_2$=0, $\beta_3$=4, $\beta_4$=0; characteristic number: $\chi$=−3.

**Utilizing $H_k$-core decomposition to simplify the process of finding highest-order cavities**

Previously, we developed some methods for computing homology, including: (1) finding shortest-length cavities via optimization [16], with advantage of obtaining optimal solutions and disadvantage of slow calculation speed; (2) finding all cavities by a spanning-tree-based algorithm [15], which may not be optimal and requires iterations. Here, a new method is introduced: (3) utilizing the $H_k$-core decomposition to find highest-order cavities. This method first performs $H_k$-core decomposition and then seek $k$-order cavities from the $H_k$-core of the original network.

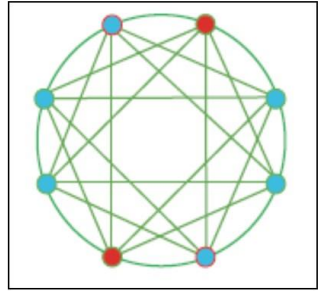

Fig. 3 A network with the smallest 3-order cavity

In [16], optimization method is applied to find, one by one, 4 shortest-length 3-order (highest-order) cavities in the C. elegans neural network: Cavity 1 {3, 13, 85, 118, 119, 158, 163, 164}; Cavity 2 {3,13, 85, 118, 119, 154, 162, 163, 164, 167, 227}; Cavity 3 {3, 13, 118, 119, 171, 173, 185, 195}; Cavity 4 {3, 13, 118, 119, 173,185, 195, 227} (see [16] for figures). Whereas, the cavity with 8 nodes is the smallest 3-order cavity, namely the totally homogeneous network shown in Fig. 3. This network has the following nested structure: 4 blue nodes constituting a square-shape cavity, which becomes a 2-order cavity with 8 triangular faces after adding the two red-circle blue nodes, and then furthermore becomes a 3-order cavity with 16 tetrahedrons after adding the two red nodes. Note also that the Cavity 2 with 11 nodes is obtained from the nested structure of a heptagon.

The new node indexes of the 16-node subnetwork of the $H_3$-core are: 3: {14, (1,0,4,0), −4}; 13: {14, (1,0,4,0), −4}; 85: {8, (1,0,2,0), −2}; 118: {14, (1,0,4,0), −4}; 119: {14, (1,0,4,0), −4}; 154: {6, (1,0,1,0), −1}; 158: {6, (1,0,1,0), −1}; 162: {6, (1,0,1,0), −1}; 163: {7, (1,0,2,0), −2}; 164: {6, (1,0,1,0), −1}; 167: {6, (1,0,1,0), −1}; 171: {6, (1,0,1,0), −1}; 173: {8, (1,0,2,0), −2}; 185: {6, (1,0,1,0), −1}; 195: {7, (1,0,2,0), −2}; 227: {8, (1,0,2,0), −2}. Here, the highest order of cavities in the node-neighbor subnetworks is 2.

Based on the nested structure, a method using the new node index is now introduced for finding the four 3-order cavities in the C. elegans neural network.

Starting from the nodes with smaller degrees, node 154 of degree 6 has neighbors 3, 13, 118, 119, 162, 167; there are no corresponding nodes with them as neighbors to form a minimum 3-order cavity. Node 158 of degree 6 has neighbors 3, 13, 85, 118, 119, 163; there is a corresponding node 164 with them as neighbors to form a minimum 3-order cavity, namely Cavity 1. The neighbors of nodes 162 and 167 of degree 6 have no corresponding nodes to form a minimum 3-order cavity. Node 171 of degree 6 has neighbors 3, 13, 118, 119, 173, 195; there is a corresponding node 185 to form a minimum 3-order cavity, namely Cavity 3. Node 195 of degree 7 has neighbors 3, 13, 118, 119, 185, 227; there is a corresponding node 173 to form a minimum 3-order cavity, namely Cavity 4. The remaining nodes, 3, 85, 118, 154, 162, 163, 164, 167, and 227, are neighbors of nodes 13 and 119, forming Cavity 2.

After ignoring some edges, the $H_3$-core consisting of 16 nodes and 68 edges is shown in Fig. 4.

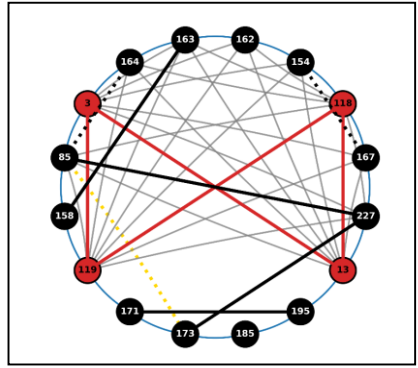

Fig. 4 The $H_3$-core of C. elegans neural network (a simplified structure)

The nodes and edges of Cavity 2 {**13**, 3, 85, 164, 163, 162, 154, 167, 227, 118, **119**} have been drawn. The remaining three cavities are as follows. Nodes 158, 171, 185 (degree 6) have 2 edges in addition to the edge connected to the red node; node 195 (degree 7) have 3 edges; node 173 (degree 8) have 4 edges, and red nodes 3, 13, 118, 119 (degree 14) should be connected to all black nodes in addition to the edge connected to the red node.

The nesting of Cavity 1 is a 1-order quadrilateral cycle (with black nodes), a 2-order cycle {3, 85, 158, 163, 164, 118}, with additional nodes 13 and 119. The nesting of Cavity 2 is a 1-order heptagonal cycle (with black nodes), a 2-order cycle {3, 85, 164, 163, 162, 154, 167, 227, 118}, with additional nodes 13 and 119. The nesting of Cavity 3 is a 1-order quadrilateral cycle (with black nodes), a 2-order cycle {13, 171, 173, 185, 195, 119}, with additional nodes 3 and 118. The nesting of Cavity 4 is a 1-order quadrilateral cycle (with black nodes), a 2-order cycle {13, 173, 185, 195, 227, 119}, with additional nodes 3 and 118.

**Example 2** Homology $H_k$-core decomposition and highest-order cavity computation for cat cortical network.

In the original cat cortical network [22], the $H_0$-core triplet ($m_k$, $r_k$, $\beta_k$) can be calculated, as follows, where the maximum nonzero $\beta_k$ has $k = 3$.

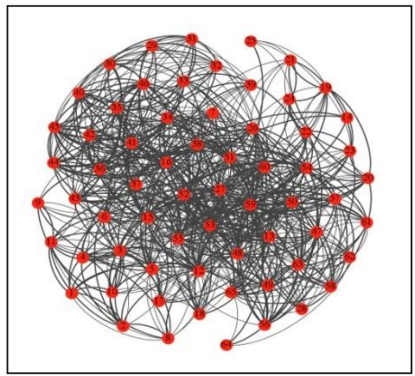

Fig. 5 Cat cortical network

Simplices: $m_0$=65, $m_1$=730, $m_2$=3613, $m_3$=10125, $m_4$=18163, $m_5$=22285, $m_6$=19290, $m_7$=11879, $m_8$=5157, $m_9$=1528, $m_{10}$=283, $m_{11}$=25; ranks of boundary matrices: $r_0$=0, $r_1$=64, $r_2$=666, $r_3$=2947, $r_4$=7176, $r_5$=10987, $r_6$=11298, $r_7$=7992, $r_8$=3887, $r_9$=1270, $r_{10}$=258, $r_{11}$=25; Betti numbers: $\beta_0$=1, $\beta_1$=0, $\beta_2$=0, $\beta_3$=2, $\beta_4$=⋯=$\beta_{11}$=0; characteristic number: $\chi$=−1.

There are two methods for finding the $H_3$-core of the cat cortical network.

**First Method:** Deleting subnetworks

In order to obtain the $H_3$-core of the network, 16 nodes with Betti numbers $(\beta_{2\text{ or }k} >0)_i$ are retained: 5, 6, 10, 12, 15, 16, 22, 23, 24, 27, 33, 41, 53, 54, 59, 60. After networking their nodes and edges, the triplet of the $H_3$-core network are checked, in which the Betti number $\beta_3$ changes from 2 to 1. From the original network of 65 nodes and edges, retaining 16 nodes and edges is equivalent to simultaneously deleting a subnetwork of 49 nodes and edges. The reason for the change in topology is that the subnetwork deletion is not performed node by node, resulting in the deletion of nodes and edges that should not be deleted being also deleted after the update.

**Second Method:** Deleting nodes

Delete nodes with a central point in the neighborhood, totally 17: 1, 2, 9, 14, 25, 29, 40, 45, 46, 55, 56, 57, 58, 61, 62, 63, 64. Then, delete nodes with Betti numbers $(1,0,0,0)_I$ totally 33: 3, 4, 5, 7, 8, 11, 13, 17, 18, 19, 21, 26, 28, 30, 31, 32, 33, 34, 35, 36, 37, 38, 39, 42, 43, 44, 47, 48, 49, 50, 51, 52, 65.

**Remark 1:** The originally retained nodes 5 and 33 become nodes with Betti numbers (1, 0, 0, 0) therefore are deleted, but the originally should be deleted node 20 with Betti numbers (1, 0, 0, 0) after the update is retained. After deletion, an $H_3$-core is obtained, with 15 nodes: 6, 10, 12, 15, 16, 20, 22, 23, 24, 27, 41, 53, 54, 59, 60.

The triplet $(m_k, r_k, \beta_k)$ of the $H_3$-core of the network are obtained as follows:

Simplices: $m_0$=15, $m_1$=67, $m_2$=129, $m_3$=116, $m_4$=49, $m_5$=12, $m_6$=1; ranks of boundary matrices: $r_0$=0, $r_1$=14, $r_2$=53, $r_3$=76, $r_4$=38, $r_5$=11, $r_6$=1; Betti numbers: $\beta_0$=1, $\beta_1$=0, $\beta_2$=0, $\beta_3$=2, $\beta_4$=$\beta_5$=$\beta_6$=0; characteristic number: $\chi$=−1.

Utilizing the method for searching highest-order cavities to find 3-order cavities. The new node index $\{n_i, (\beta_k)_i, \chi_i\}$ of $H_3$-core subnetwork with 15 nodes are: 12: {11, (1,0,1), −1}, 27: {11, (1,0,1), −1}, 53: {11, (1,0,1), −1}, 60: {11, (1,0,1), −1}; 16: {10, (1,0,1), −1}, 54: {10, (1,0,1), −1}; 10: {8, (1,0,1), −1}, 15: {8, (1,0,1), −1}; 22: {8, (1,0,1), −1}, 23: {8, (1,0,1), −1}; 6: {7, (1,0,1), −1}, 20: {7, (1,0,1), −1}; 24: {6, (1,0,1), −1}, 41: {6, (1,0,1), −1}; 59: {12, (1,0,2), −2}. Here, the highest order of cavities in node-neighbor subnetworks is 2.

Next, the above indices are used to find the two 3-order cavities in the network. Starting from the nodes with small degrees, node 41 with degree 6 has neighbors: 6, 10, 12, 15, 16, 53; and the corresponding node 59 forms a minimum 3-order cavity with them as neighbors, namely Cavity 1 {41, 6, 10, 12, 15, 16, 53, 59}. Node 24 with degree 6 has neighbors: 20, 22, 23, 27, 54, 60; and the corresponding node 59 forms another minimum 3-order cavity, namely Cavity 2 {24, 20, 22, 23, 27, 54, 60, 59}. The subnetwork of node 59 has two 2-order cavities, which form two 3-order cavities with the corresponding nodes 24 and 41, respectively.

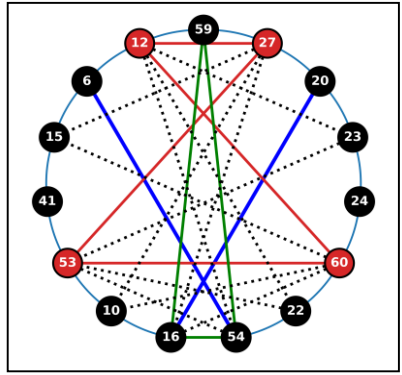

Fig. 6 The $H_3$-core of cat cordial network (a simplified structure)

The nodes with degree 6 are 24 and 41, each of which can only form a 3-order cavity. The nodes with degree 7 are 6 and 20, each of which can form a 3-order cavity and they have 2 blue cross-cavity edges. The nodes with degree 8 are 10, 15, 22 and 23, each of which can form a 3-order cavity and they have 8 edges connecting with the cross-cavity red nodes. The nodes with degree 10 are 16 and 54, each of which can form a 3-order cavity and they have 3 cross-cavity edges (two blues and one green) and 4 edges connecting with the cross-cavity red nodes. The red nodes with degree 11 are 12, 27, 53 and 60, each of which can form a 3-order cavity and they have 4 edges connecting with the cross-cavity red nodes.

**Remark 2:** The algorithm first deletes central nodes, and then compute Betti numbers until order $k$, which can reduce the computational burden.

## 4 Discussion

Some interesting and closely related issues are discussed in this section.

**Quickly determining the $H_k$-core ($k≥3$) of a given network**

The highest-order cavities in a network usually have a symmetrical structure in the deep layers of the network. Although the number of these cavities are generally not large, but they are hidden in the highest-order $H_k$-core. How to quickly determine the $H_k$-core ($k≥3$) for a network with a large number of nodes? The two methods introduced in **the Results section** are sometimes not very effective. In order to obtain the $H_k$-core as soon as possible, it is recommended to use nodes from the $k$-order cavity-generating cliques [16] and nodes from $\beta_{k-1}$ branches in the subnetworks of those clique nodes to form a network together. Thus after simplification, the $H_k$-core can be obtained.

For example, the cavity-generating cliques of four 3-order cavities in the C. elegans neural network [16] are: (118, 119, 163, 164), (118, 119, 167, 227), (118, 119, 185, 195), and (118, 119, 195, 227). Whereas the 2-order cavities in the subnetworks with different nodes are as follows:
Node 118: {13,85,119,158,163,164}, {13,87,119,158,163,164}, {13,119,171,173,185,195},
 {13,119,173,185,195,227}, {85,87,120,143,158,192}, {87,88,110,120,143,192},
 {13,85,119,154,162,163,164,167,227};
Node 119: {3,85,118,158,163,164}, {3,118,171,173,185,195}, {3,118,173,185,195,227},
 {3,85,118,154,162,163,164,167,227};

Node 163: {3,13,118,119,158,164}, {3,13,118,119,162,164};
Node 164: {3,13,85,118,119,163};
Node 167: {3,13,118,119,154,227};
Node 185: {3,13,118,119,173,195};
Node 195: {3,13,118,119,171,185}, {3,13,118,119,185,227};
Node 227: {3,13,85,118,119,167}, {3,13,85,118,119,195}, {3,13,118,119,177,195}.
Therefore, the nodes that make up the $H_3$-core network are: 3, 13, 85, 87, 88, 110, 118, 119, 120, 143, 154, 158, 162, 163, 164, 167, 171, 173, 177, 185, 192, 195, 227. Delete newly appeared nodes: 87, 88, 110, 120, 143, 177, 192 with Betti numbers $(1, 0, 0, 0)_i$ of the network. Finally, the $H_3$-core network is obtained.

**Potential applications of the $H_k$-core decomposition**

This paper introduces network homology $H_k$-core decomposition, focusing on its application in simplifying the search for network higher-order cavities. In fact, $H_k$-core decomposition has many potential applications in network sorting and higher-order dynamics.

After performing the network $H_k$-core decomposition, the $H_k$-core of the network nodes can be obtained according to the higher-order core of the nodes and the Betti numbers of the neighboring subnetwork. For the C. elegans neural network, the nodes 3, 13, 118, and 119 are ranked first in the $H_k$-core because their higher-order core is the highest $H_3$ and their Betti number is the largest, $\beta_2=4$.

With the $H_k$-core decomposition of the network, the $H_k$-core percolation of the network can be discussed by referring to the network interdependent percolation and the $k$-core percolation. It is well known that the network interdependent percolation is a continuous phase transition at $l<4$, and a mixed phase transition will occur at $l\geq 4$, where $l$ is the constraint distance of the dependent node pair [24]. The $k$-core percolation is a continuous phase transition at $k<3$, and a mixed phase transition will occur at $k\geq 3$ [24]. Therefore, it is reasonable to speculate that $H_k$-core percolation is a continuous phase transition at $k<2$, and a mixed phase transition will also occur at $k\geq 2$.

By using the $H_k$-core sorting of nodes, one can also study the most influential propagation sources of higher-order networks [25], the optimal pinning control simplices of higher-order networks [26], and so on.

**Utilizing $H_k$-shell decomposition to simplify cavity search**

**Example 3** Consider the 1-oreder cavity of the sample network shown in Fig. 2.

Referring to the idea of $H_k$-core decomposition of network homology, as long as the $H_2$-core is deleted from the $H_1$-core, an $H_1$-shell can be obtained [27]. The method is as follows: For node $i$ in $(\beta_1>0)_i$, delete the edge connecting with $\beta_1$ branche in the neighbor subnetwork. Then, in order to reduce the value of $\beta_2$, it is not just to delete the edge, but actually delete the triangle. In the $H_1$-core network, select node 9 to delete the edge 9-13, which connects the subnetwork to the 1-order cavity. In fact, two triangles are deleted at the same time: ▲9-10-13 and ▲9-12-13, so the network characteristic number is reduced by 1. After updating, delete nodes 13, 12, and 11 one by one, till the process ends.

The final $H_1$-shell has the triplet $(m_k, r_k, \beta_k)$ as follows:

Simplices: $m_0=9$, $m_1=10$; ranks of boundary matrices: $r_0=0$, $r_1=8$; Betti numbers: $\beta_0=1$, $\beta_1=2$; characteristic number: $\chi=-1$. The two shortest-length 1-order cavities are: Cavity 1: {3, 6, 7, 8} and Cavity 2: {1, 5, 9, 10, 14, 6, 3}.

**Example 4** The 2-order cavities of the C. elegans neural network.

The method for deleting the $H_3$-core in the $H_2$-core of the C. elegans neural network is as follows: For node $i$ of $(\beta_2>0)_i$, delete the edge connecting with $\beta_2$ branche in the neighbor subnetwork, select nodes 154 (degree 8), 185 (degree 9), 158 (degree 10), 171 (degree 10) in the $H_2$-core, and delete the edges 154-118, 185-195, 158-3, 171-195 connecting the subnetwork with the 2-order cavity. Then, the $H_2$-shell network (152, 1165) is obtained. Deleting an edge here actually deletes four triangles and four tetrahedrons at the same time, so the network characteristic number increases by 1.

The final $H_2$- shell network has the triplet ($m_k$, $r_k$, $\beta_k$) as follows:

Simplices: $m_0=152$, $m_1=1164$, $m_2=1996$, $m_3=1261$, $m_4=556$, $m_5=191$, $m_6=36$, $m_7=2$; ranks of boundary matrices: $r_0=0$, $r_1=151$, $r_2=1013$, $r_3=862$, $r_4=399$, $r_5=157$, $r_6=34$, $r_7=2$; Betti numbers: $\beta_0=1$, $\beta_1=0$, $\beta_2=121$, $\beta_3=0$, $\beta_4=\cdots=\beta_7=0$; characteristic number: $\chi=122$.

The search for 2-order cavities in the C. elegans neural network in the $H_2$-shell reduces the search size from 3241 triangles to 1013 triangles compared to the original network.

**Introduction to Network Homology Calculations**

The above has discussed the homology $H_k$-core decomposition defined by boundary operators. The $H_k$-core decompositions of other homologies, such as the $H_k$-core decompositions of cohomology and path homology, can be similarly discussed.

The boundary operator $\partial_k$: $C_k \rightarrow C_{k-1}$ of the homology is a mapping from higher dimension to lower dimension, and the exterior operator of the cohomology [18,19] $\delta_{k-1}$: $C^{k-1} \rightarrow C^k$ is a mapping from lower dimension to higher dimension. The exterior matrix of the exterior operator $\delta_{k-1}$ is the transposed matrix of the boundary matrix of the boundary operator $\delta_k$ [19]. The homology group on the network is $H^k = \ker(\delta_k)/\mathrm{im}(\delta_{k-1})$ [19].

Path homology is related to directed networks. The edges of a directed network, namely, the 1-order simplices, are directed, and they point from the source (starting node) to the sink (ending node). How do higher-order simplices determine their directions? They also point from the source to the sink. For example, the three nodes 1, 2, and 3 of the 2-order simplex have three directed edges connecting them: 2→3, 2→1, 3→1. This means that 2 is the source node and it points to the sink node 1. The writing order of the 2-order directed simplex 231 reflects the directed path 2→3→1. The 3-order directed simplex 1324 not only shows its directed path 1→3→2→4, but also determines six directed edges. However, the three directed edges 1→2, 2→3, 3→1 have no source node or sink node, and do not form a 2-order directed simplex; it is just a 1-order cycle. The importance of paths to directed networks is self-evident, and path homology [28] came into being.

Persistent homology is related to weighted networks. In addition, as the distance decreases, the nested simplicial complex in the filter increases. However, as the number of nodes and edges increases or decreases in the temporal network, the simplicial network at different times can be large or small, and they have no nested relationship. Therefore, the mapping direction between networks can be forward or backward, which involves zigzag persistent homology [29].

As a super network of higher-order networks, various homologies are also being introduced to study higher-order structures [30].

**Platonic Polyhedron Networks**

The Platonic polyhedrons [31] are named after the research done by Plato and his followers. They are regular polyhedrons in three-dimensional Euclidean space that satisfy the conditions of uniform node degrees, consistent edge lengths, identical vertex angles, and congruent faces. Such highly symmetric geometric bodies, when discretized, have a close relationship with higher-order network homology groups.

If the nodes, edges, and triangles of Platonic polyhedrons are considered as simplices of higher-order networks, then discrete Platonic polyhedron networks can be obtained from continuous regular polyhedrons. The five types of Platonic polyhedron networks are as follows.

Tetrahedron: 4 nodes, 6 edges, 4 triangles; it is a 3-order simplex;

Octahedron: 6 nodes, 12 edges, 8 triangles; it is a minimum 2-order cavity;

Icosahedron: 12 nodes, 30 edges, 20 triangles; it is a 2-order cavity;

Hexahedron: 8 nodes, 12 edges, 6 quadrilaterals; if each face adds 1 diagonal edge, then it can form a 2-order cavity;

Dodecahedron: 20 nodes, 30 edges, 12 pentagons; if each face adds 1 node and 5 edges, then it also forms a 2-order cavity.

Three-dimensional regular polyhedrons are generalizations of two-dimensional regular polygons. Further, higher-dimensional regular polytopes can also be defined. There are infinitely many polytopes in two dimensions, five in three dimensions, six in four dimensions, and three in five dimensions and above. This result was first derived by Schläfli and proved by Coxeter in 1969 [31].

First, except for triangles, all other two-dimensional regular polygons may be 1-order cavities. Then, by adding nodes and edges to form nested layers and splicing combination of triangulation, infinitely many higher-order cavities can be generated. Although the cavities obtained in this way will lose some symmetry, an interesting question is whether these configurations can include all cavities?

The 2-order cavities in the C. elegans neural network include the **octahedrons**, for example the cavity {3, 4, 13, 64, 87, 102}; the **hexahedron** has one diagonal edge on each face, for example the cavity {1, 3, 4, 6, 9, 13, 64, 202}, etc. The 2-order cavities of the **icosahedron** and the 2-order cavities of the **dodecahedron** with a node and five edges on each face are expected to be found in more complex networks.

**References**
1. Watts D, Strogatz S. Collective dynamics of 'small-world' networks. *Nature*, 1998; 393, 440-442.
2. Barabási A-L, Albert R. Emergence of scaling in random networks. *Science*, 1999, 286(5439), 509-512.
3. Wang X, Chen G. Synchronization in scale-free dynamical networks: Robustness and fragility. *IEEE Trans Circuits Syst.* I, 2002; 49, 54-62.
4. Shi D, Chen G, Thong WWK, Yan X. Searching for optimal network topology with best possible synchronizability. *IEEE Circuits Syst. Mag.,* 2013, 13, 66-75.
5. Shi D, Lü L, Chen G. Totally homogeneous networks. *Natl Sci Rev*. 2019; 6(5):962–969.
6. Shi D, Chen G. Simplicial networks: A powerful tool for characterizing higher-order interactions. *Natl Sci Rev.*, 2022; 9(5), nwac038.
7a. Fan T, Lü L, Shi D. Towards the cycles structure in complex network: A new perspective. arXiv: 1903.01397
7b. Fan T, Lü L, Shi D, Zhou T. Characterizing cycle structure in complex networks. *Commun Phys*. 2021, 4(272), 1-9.
8. Battiston F, Cencetti G, Iacopini I, Latora V, Lucas M, Patania A, Young J-G, Petri G. Networks beyond pairwise interactions: Structure and dynamics. *Phys Rep*., 2020, 874, 1-92.
9. Hatcher A. *Algebraic topology*. Cambridge (UK): Cambridge University Press, 2002.
10. Tymochko S, Munch E, Khasawneh, F-A. Using zigzag persistent homology to detect Hopf bifurcations in dynamical systems. *Algorithms*, 2020, 13, 278.
11. Morley-Fletcher, R. Big data: What is it and why is it important? In: *Digital agenda for Europe.* European Commission. TechTarget/Data Management, online; 2013.
12. Carlsson G. Topology and data. *Bull Am Math Soc*. 2009, 46, 255-308.
13a. Edelsbrunner H, Harer J. Persistent homology—A survey. *Contemp Math*., 2008, 453, 257-282.
13b. Edelsbrunner H, Letscher D, Zomorodian A. Topological persistence and simplification. *Discrete Comput Geom*., 2002, 28, 511-533.
14. Zomorodian A, Carlsson G. Computing persistent homology. *Discrete Comp. Geom.* 2005, 33, 249-274.
15. Shi D, Chen Z, Ma C, Chen G. Computing persistent homology by spanning trees and critical simplices. *Research A Science Partner Journal* 2023; 6: 0230.
16. Shi D, Chen Z, Sun X, Chen Q, Ma C, Lou Y, Chen G. Computing cliques and cavities in networks. *Comm. Phys*. 2021, 4, 249.
17. Battiston F, Petri G. *Higher-order systems. Complexity*. Berlin: Springer; 2022.
18. Dey T, Wang Y. *Computational topology for data analysis*. Cambridge: Cambridge University Press, 2022.
19. Bianconi, G. *Higher-order networks: An introduction to simplicial complexes*. Cambridge (UK): Cambridge University Press, 2020.
20. Gu, X F, Yau, S T. *Computational conformal geometry*: *Theory*. International Press of Boston, Inc., 2008.
21. Lim L-H, Hodge Laplacians on graphs. *SIAM Review*, 2020, 62(3), 685-715.
22. Rossi R A, Ahmed N K. The network data repository with interactive graph analytics and visualization. In: *29th AAAI Conference on Artificial Intelligence*, 2015, 4292-4293.
23. Kitsak M, Gallos L K, Havlin S, Liljeros F, Muchnik L, Stanley H E, Makse H A. Identification of influential spreaders in complex networks. *Nature Phys.,* 2010, 6, 888-893.

24. Gao S, Xue L, *et al*. Possible origin for the similar phase transitions in *k*-core and interdependent networks. *New J. Phys., 2024,* 26, 013006.
25. Hu Y Q, Havlin S, *et al.* Local structure can identify and quantify influential global spreaders in large scale social networks, *PNAS, 2018,* 3, 201710547.
26. Zhou J, Li B, Lu J, Shi D. Selection of simplexes in pinning control of higher-order networks. *SCIENTIA SINICA* Informatics, 2023, 23.
27. Carmi S, Havlin S, Kirkpatrick S, Shavitt Y, Shir E. A model of Internet topology using k-shell decomposition. *PNAS* 2007, 104, 11150-11154.
28. Grigor'yan A, Lin Y, Muranov Y, Yau S-T, Homologies of path complexes and digraphs, *Pure and Appl. Math. Quarterly,* 2014, 10(4), 619-674.
29. Carlsson G, de Silva V. Zigzag persistence. *Found. Comp. Math.,* 2010, 10(4), 367-405.
30. Gasparovic E, Wang B, Ziegelmeier A, *et al.* A survey of simplicial, relative, and chain complex homology theories for hypergraphs. arXiv: 2049.18310v1, 2024.
31. Coxeter H S M. *Introduction to geometry*, Wiley, 2022.